\documentclass[12pt,titlepage]{article}

\usepackage[parfill]{parskip}    \usepackage[english]{babel}
\usepackage[applemac]{inputenc}

\usepackage{fullpage}
\usepackage{hyperref}
\usepackage{graphicx}
\usepackage{amssymb}
\usepackage[scale=0.8]{geometry}

\usepackage{multirow}
\usepackage[round,numbers,sort&compress]{natbib} 

\usepackage{setspace}

\begin{document}

\newpage

\textbf{\large{Photoluminescent diamond nanoparticles for cell labeling: study of the uptake mechanism in mammalian cells}}

\emph{Orestis Faklaris $^\star$, Vandana Joshi $^\dagger$, Theano Irinopoulou~$^\P$, Patrick Tauc $^\S$, Hugues Girard $^\ddagger$, Celine Gesset $^\ddagger$, Mohamed Senour $^\flat$, Alain Thorel $^\flat$, Jean-Charles Arnault $^\ddagger$, Jean-Paul Boudou$^\dagger$, Patrick A. Curmi $^\dagger$*, and Fran\c cois Treussart  $^\star$*} \\

$^\star$ O.~Faklaris, Prof.~F.~Treussart \\
Laboratoire de Photonique Quantique et Mol\'eculaire, UMR CNRS 8537, Cachan (France)\\

 $^\dagger$ Dr. V.~Joshi, Dr.~J.-P.~Boudou, Dr. P.A.~Curmi\\
 Laboratoire Structure et Activit\'e des Biomol\'ecules Normales et Pathologiques, Universit\'e Evry-Val-d'Essonne and INSERM U829, Evry (France)\\
 
  $^\P$ Dr. T.~Irinopoulou\\
 Institut du Fer à Moulin, UPMC Univ. Paris 6 et INSERM, UMR-S 839  (France)\\
 
 $^\S$ Dr. Patrick Tauc\\
	 Laboratoire de Biologie et de Pharmacologie Appliquée,\\
	  Ecole Normale Supérieure de Cachan, CNRS UMR 8113, Cachan, (France)\\

  $^\ddagger$Dr. H.~Girard, Dr. C.~Gesset, Dr. J.C. Arnault\\
  Diamond Sensor Laboratory, \\
	  Centre d'Etudes Atomiques, Gif-sur-Yvette, France

  $^\flat$Dr. M.~Sennour, Dr. A.~Thorel,\\
    Laboratoire Pierre-Marie Fourt CNRS UMR 7633,\\
   Centre des Matériaux de l'Ecole des Mines de Paris, Evry (France)\\
   
 [*] corresponding author:  F. Treussart and P. A. Curmi\\
 francois.treussart@ens-cachan.fr and pcurmi@univ-evry.fr \\

\clearpage

\begin{abstract}
Diamond nanoparticles (nanodiamonds) have been recently proposed as new labels for cellular imaging. For small nanodiamonds (size $<40~$nm) resonant laser scattering and Raman scattering cross-sections are too small to allow single nanoparticle observation. 
Nanodiamonds can however be rendered photoluminescent with a perfect photostability at room temperature. Such a remarkable property allows easier single-particle tracking over long time-scales.
In this work we use photoluminescent  nanodiamonds of size $<50$~nm for intracellular labeling and investigate the mechanism of their uptake by living cells . By blocking selectively different uptake processes we show that nanodiamonds enter cells mainly by endocytosis and converging data indicate that it is clathrin mediated. 
We also examine nanodiamonds intracellular localization in endocytic vesicles using immunofluorescence and transmission electron microscopy. We find a high degree of colocalization between vesicles and
 the biggest nanoparticles or aggregates, while the smallest particles appear free in the cytosol. Our results pave the way for the use of photoluminescent nanodiamonds in targeted intracellular labeling or biomolecule delivery.\


\bigskip

\end{abstract}

\section{Introduction}
Solid-state nanoparticles hold great promises for biomedical applications, notably thanks to the possibility to combine biological and inorganic materials with the prospect to develop innovative diagnostic and therapeutic tools. Among them, nanoparticles like quantum dots~\cite{Alivisatos04, Michalet_Science05}, gold nanobeads~\cite{Lounis06} or silicon beads~\cite{Tilley05} are used to label biomolecule with high specificity, to track their fate in cultured cells and in organisms or even to deliver bioactive molecules or drugs.
Organic dyes and fluorophores are nowadays the most widely used biomolecules fluorescent labels. However they photobleach rapidly under continuous illumination~\cite{Nann08}, which makes difficult their quantification and long term 
follow-up. Interestingly, semiconductor nanocrystals, or quantum dots (QDs), have a better stability and a lower photobleaching yield than organic dyes. They also offer the possibility of multicolor staining by size tuning~\cite{Michalet_Science05}. On the other hand, such nanoparticles present major drawbacks, such as a potential cytotoxicity on long-term scale due to the chemical composition of their core~\cite{Parak_cytotox_05} or the intermittency of their photoluminescence (photoblinking) which makes difficult an efficient tracking of individual nanoparticles~\cite{Nirmal_96}.  

Compared to these nanoparticles, photoluminescent nanodiamonds (PNDs) appear as promising alternative biomarkers. Their photoluminescence, with emission in the red and near infrared spectral region (575-750~nm) results from nitrogen-vacancy (NV) color centers embedded in the diamond matrix~\cite{Gruber_97}. These emitters present a perfect photostability with no photobleaching nor photoblinking, and a photoluminescence intensity which is only three times smaller than that of a single commercial QD, a situation which can be reversed in favor of NDs by the use of particles containing a large number of color centers~\cite{Faklaris09}.
Such remarkable photoluminescent properties have made possible long-term single particle tracking in living cells~\cite{Neugart_07,Chang_08, Faklaris09}.

Moreover, thanks to a large specific surface area, the nanodiamond can be used as a platform to graft a large variety of bioactive moieties~\cite{Kruger_NDs_2008, Kruger09, Cheng_Biophys_J, Vial08} with the perspective to use these particles for instance as intracellular compartment labels~\cite{Opitz09} or as long term traceable delivery vehicles for biomolecule translocation in cell~\cite{Cheng_Biophys_J, Vial08, Huang07, Osawa09}.
Considering the recent demonstrations of PNDs mass production~\cite{Curmi09, Chang_08}, and the numerous  experiments showing their low cytotoxicity~\cite{Fu07, Faklaris_08, Schrand_07, Huang07}, one can envision large scale applications of these biolabels. However, such outstanding prospects require a better understanding of the mechanism underlying their cellular uptake.

To that end, we study in the present work the cellular fate of photoluminescent nanodiamonds of size $<$50~nm. We first show that for such a small size nanodiamonds, excitation-laser scattering yields no signal and that only photoluminescence allows their detection with a sufficient signal-to-background ratio. We then carry out a systematic investigation of the cellular uptake and pathway of PNDs in human cancer cells (HeLa cells) and observe that the internalization of PNDs stems mainly from endocytosis.

The localization of PNDs in cells is studied using simultaneous detection of PNDs and immunofluorescence methods to label the endosomes and the lysosomes. 

\section{Results and discussion}
\subsection{Nanodiamond observation at the single particle level in cells}

In the present work, we use nanodiamonds produced by milling of 150-200~$\mu$m diamond microcrystals after the activation of their photoluminescence by the creation of NV color centers in the diamond matrix (see Materials and Methods). We then selected from an aqueous suspension of the milled product, using differential centrifugations, a subset of nanodiamonds having a mean hydrodynamic diameter centered on 46~nm. Since diamond refractive index (2.4) is about twice larger than that of the cellular medium, diamond can yield a higher back-scattered intensity of the excitation laser light than the cell compartments. Such a scattering signal can then be used to image nanodiamonds in cell with a good contrast~\cite{Neugart_07, Plakhotnik_liposomes}. However, the scattering intensity decreases rapidly with particle size, due to its sixth order dependence on the particle diameter (Rayleigh scattering). It was shown that the smallest nanodiamonds detectable by this technique have sizes of 37~nm~\cite{Plakhotnik_OptLett06}.
\begin{figure}[ht!]
\begin{center}
\includegraphics[width=0.75\textwidth]{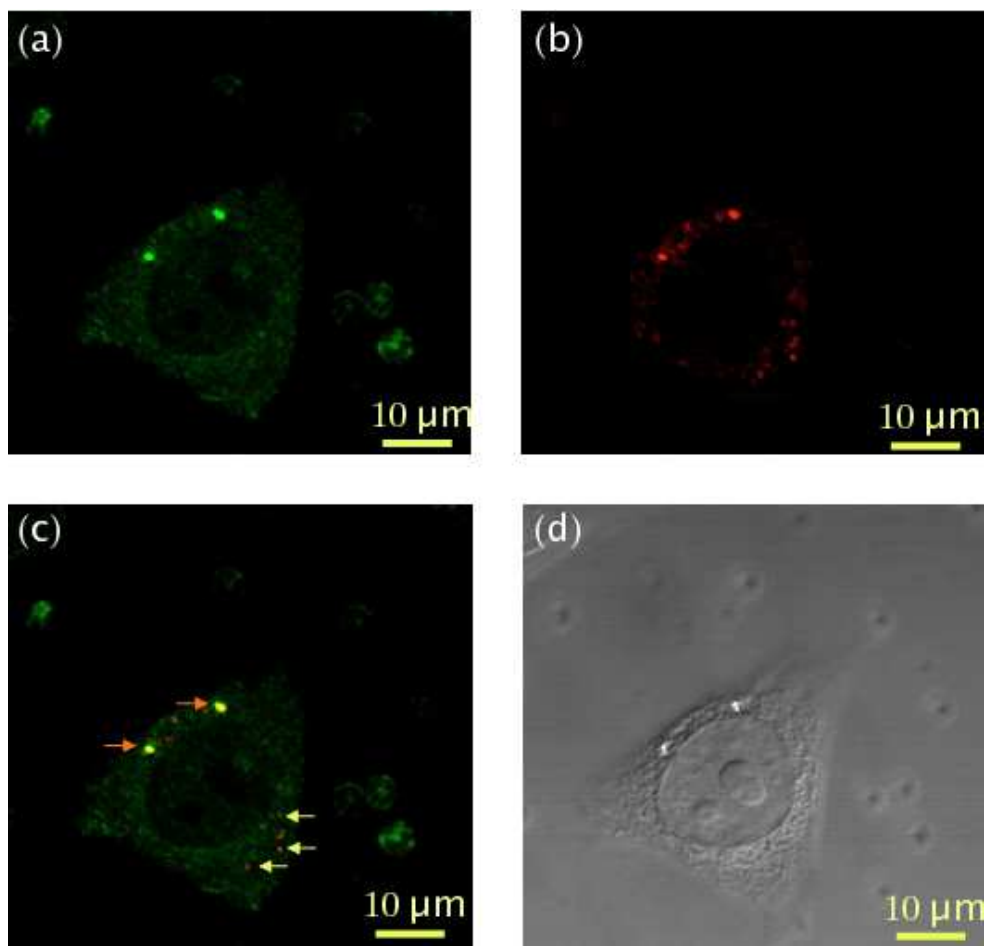}
\caption{Imaging of photoluminescent nanodiamonds internalized in HeLa cells in three different contrast modes of a Leica TCS SP2 microscope: (a) raster-scan of back-scattered excitation laser light (wavelength 488~nm). (b) photoluminescence confocal raster-scan in the NV color center emission spectrum region (\emph{red channel}, refer to Materials and Methods). (c) overlay of (a) and (b): orange arrows, pointing to the right, exhibit nanoparticles that are observed in both backscattered and fluorescence modes, while the yellow arrows pointing to the left show some nanodiamonds that are detected only in the photoluminescence mode; (d) DIC image obtained in transmission with white light illumination. The confocal section (b) was acquired with laser focusing into the plane located $2.5~\mu$m above the coverglass surface on which the cell was grown.}
\label{fig:figure1}
\end{center}
\end{figure}

Accordingly, we observe that NDs size is too small to image the particles in cell by back-scattered light with a  reasonable signal-to-background ratio that can compete with the photoluminescence signal. 
{Figure~\ref{fig:figure1}} shows a HeLa cell imaged with a standard confocal microscope (Leica TCS SP2) in three different contrast modes: (a) backscattering (sometimes referred as the ``reflection'' mode) of the incident laser operating in cw regime at an excitation wavelength of 488~nm, (b) photoluminescence excited by the same laser and detected in the NV color center spectral region and (d) DIC (Differential Interference Contrast) obtained in transmission under white light illumination of the sample.  Two PNDs are detected 
by all the contrast modes and probably correspond to particles bigger than 50~nm (orange arrows pointing to the right on {Figure~\ref{fig:figure1}c}). On the contrary, a few PNDs are only detected in the fluorescence mode (three of them are marked with yellow arrows pointing to the left on {Figure~\ref{fig:figure1}c}). Complementary observations done with a home-made confocal microscope having single color-center sensitivity confirm this result. 
These experiments indicate that ﬂuorescence microscopy is the most appropriate method to detect the smallest PNDs internalized by cells.
Moreover, a higher PNDs photoluminescence intensity can be achieved by increasing the number of color centers created per particle.
A recent report on PNDs produced a technique similar to the one used here shows that, by optimizing the NV color center creation, one can reach a concentration up to 12~NV centers in nanodiamonds of size around 10~nm~\cite{Curmi09}. Such a result is very promising for the tracking of very small PNDs in cell.

\subsection{Mechanism of PNDs uptake by cells}
It is well established that NDs can spontaneously enter cultured cells, as demonstrated by confocal microscopy~\cite{Fu07, Faklaris_08, Cheng_Biophys_J}. In this work we further investigated the internalization mechanism of PNDs in HeLa cells.

We quantified the dynamics of the PNDs uptake by cells by monitoring the overall cell fluorescence intensity (refer to Materials and methods) and we found that the uptake follows an exponential behavior with time, with a characteristic uptake half-life of 2.6~h. These measurements are in agreement with the 3~h uptake half-life observed for 70~nm PNDs~\cite{HCChang_DiamRelMat09} and the 1.9~h reported for 50~nm gold nanoparticles~\cite{Chithrani06}.

Endocytosis is considered as the dominant uptake mechanism of extracellular materials of size up to about 150~nm~\cite{Marsh99, Murkherjee97}.
We therefore evaluated the contribution of endocytosis to NDs internalization by HeLa cells. For this purpose, the cells were incubated with PNDs (concentration 20~$\mu$g/ml) under different conditions: (i) at 37$^\circ$C (control), (ii) at 4$^\circ$C and (iii) after pretreatment with NaN$_3$. The latter treatment disturbs the production of ATP and blocks the endocytosis~\cite{Schmid90} which is an energy-dependant process. Incubation of the cells at 4$^\circ$C is also known to block endocytosis.

\begin{figure}[ht!]
\begin{center}
\includegraphics[width=0.9\textwidth]{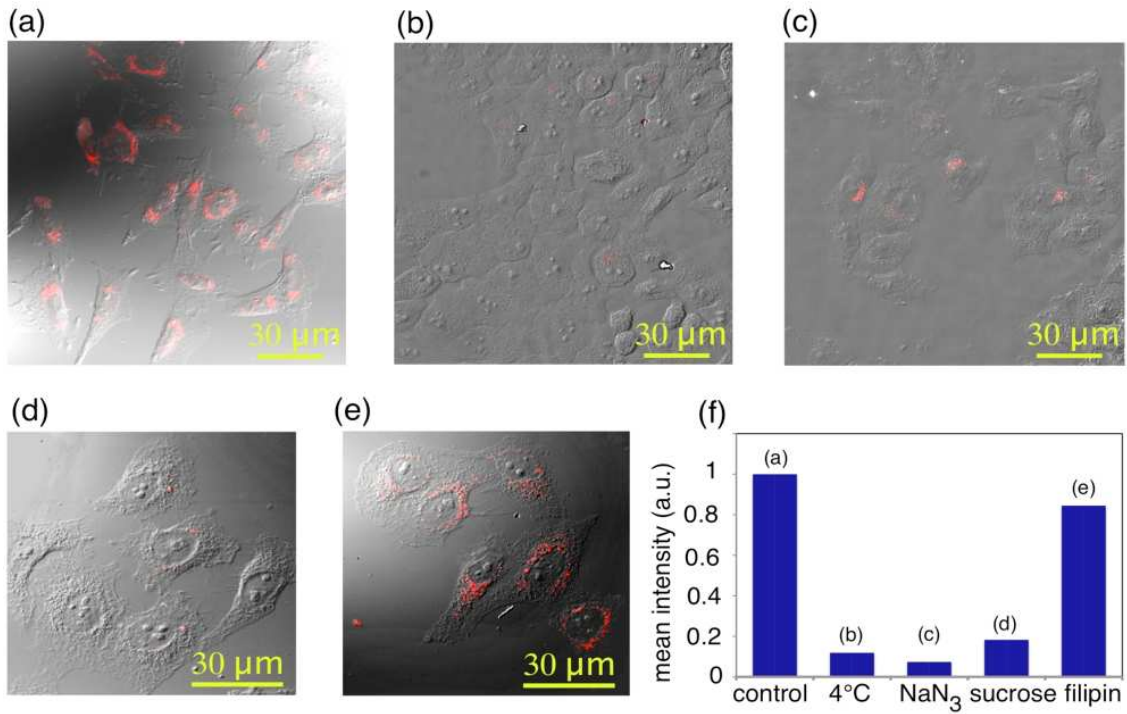}
\caption{Nanodiamonds are uptaken by HeLa cells through endocytosis. Merged photoluminescence confocal raster scans (\emph{red channel}) and DIC images of PNDs (concentration 20~$\mu$g/ml) incubated for 2~hours with cells (a) at 37$^\circ$C (control) and (b) at 4$^\circ$C, or at 37$^\circ$C but after pretreatment with either (c) NaN$_3$ (10~mM), or (d) sucrose (0.45~M), or (e) filipin (5~$\mu$g/ml). Confocal scans are acquired at $z=1.5~\mu$m above the coverglass surface, laser excitation power: 0.5~mW. (f) Mean photoluminescence intensity per cell (in the \emph{red channel}) for the different cell treatments, normalized to the one of control cells and evaluated as described in Materials and Methods.}
\label{fig:figure2}
\end{center}
\end{figure}

For the analysis of NDs uptake under different cellular treatments, we used the Leica TCS SP2 confocal microscope. {Figure~\ref{fig:figure2}} shows that when endocytosis is hindered by low temperature or NaN$_3$, the photoluminescence signal from PNDs strongly decreases ({Figure~\ref{fig:figure2}b,c}) compared to the control ({Figure~\ref{fig:figure2}a}). Similar results were obtained using the home-built confocal microscope which is able to detect all PNDs, including those containing only a single color center, presumably the smallest nanoparticles.
These observations, together with the fact that the majority of the NDs from the sample exhibit some photoluminescence ({Figure~\ref{fig:figure5}}, Materials and methods), strongly suggest that very few NDs are internalized when endocytosis is blocked.

To further documant the endocytosis mechanisms involved in the cellular uptake of nanodiamonds, we investigated the receptor-mediated endocytosis (RME) pathway. In RME a ligand first binds to a cell surface receptor and is then internalized through an invagination of the plasma membrane. Among the different RME processes, the clathrin mediated pathway is the most frequent one. Clathrin is a protein which coats cell membrane invaginations leading to the budding of clathrin-coated vesicles~\cite{Steinman83, Mousavi_Clathrin_04}. Other main RME processes, which are clathrin-independent, occur through the caveolae pathway. Caveaolae are invaginations rich in cholesterol~\cite{Pelkmans04}.
In our experiments, cells were incubated with PNDs under conditions that inhibit either the clathrin or the caveolae pathways.
Interestingly, we observe that pretreatment of cells with sucrose, a hypertonic treatment known to disrupt the formation of clathrin-coated vesicles~\cite{Anderson89, Qaddoumi03}, reduces to a high degree the PNDs uptake ({Figure~\ref{fig:figure2}d}).
To block the caveolae pathway, cells were pretreated with filipin which disrupts the formation of the cholesterol domains~\cite{Kruth_filipin_80, Qaddoumi03}. In contrast to the clathrin-pathway blocking experiment, we observe that pretreatment with filipin does not hinder the internalization of PNDs ({Figure~\ref{fig:figure2}e}). These results indicate that PNDs are mainly internalized by cells by the clathrin-mediated pathway. 

For a more quantitative analysis, we evaluated the mean photoluminescence intensity per cell (in the \emph{red channel}, corresponding to NV color center emission), in a way similar to the one used for the dynamic measurement of PNDs cellular internalization.
{Figure~\ref{fig:figure2}f} shows the change of the mean photoluminescence intensity per cell, normalized to that  of control cells. This graph summarizes quantitatively the effects of the different cell treatments, and supports the conclusion that the uptake mechanism of PNDs is endocytosis, with strong indications that it is clathrin mediated.
The latter statement is reinforced by a complementary analysis done on nanodiamond surface functions. We studied by FTIR spectroscopy the modifications of ND surface functions signatures before and after their mixture with serum supplemented culture medium. We observed that proteins of the serum are adsorbed onto the nanodiamond surface, which may facilitate a receptor-mediated endocytosis. Concomitantly, we noticed that the addition of serum greatly improves the stability of the nanodiamond suspension in the culture medium, which is also related to surface modifications impacting its electric charge. 

The present results on nanodiamond internalization pathway are close to those obtained on the same HeLa cell line for other types of nanoparticles with similar sizes, like gold nanoparticles~\cite{Chithrani06} or single-walled carbon nanotubes noncovalently conjugated with DNA molecules or proteins~\cite{Dai06}. 

\subsection{Intracellular localization of PNDs}
After an endocytic uptake, the internalized compound is expected to be found in intracellular endosomal and lysosomal vesicles, before eventually being released in the cytosol or expelled from the cell.
Endosomes are vesicles involved in the transport of extracellular materials in the cell cytoplasm. After internalization, endosomes are either recycled to the plasma membrane and the receptors can be used for a novel cycle or they are fused with lysosomes~\cite{Steinman83}.

In a previous study we concluded from immunofluorescence analyses that 25~nm PNDs are partially colocalized with early endosomes~\cite{Faklaris_08}. Here we provide similar but more complete results obtained with PNDs differently produced. Apart from a difference in the mean size between the two kinds of PNDs, the nanoparticles have not the same morphologies as it can be seen on TEM images (see {Figure~\ref{fig:figure5}c} in the Materials and Methods and also Ref.\cite{Curmi09}): the shape of the 46~nm nanoparticles produced by milling is more spherical, while the 25~nm commercial NDs present sharper edges~\cite{Treussart_PhysB06}. These differences may impact the uptake efficiency and justify a new localization study.

\begin{figure}[ht!]
\begin{center}
\includegraphics[width=0.8\textwidth]{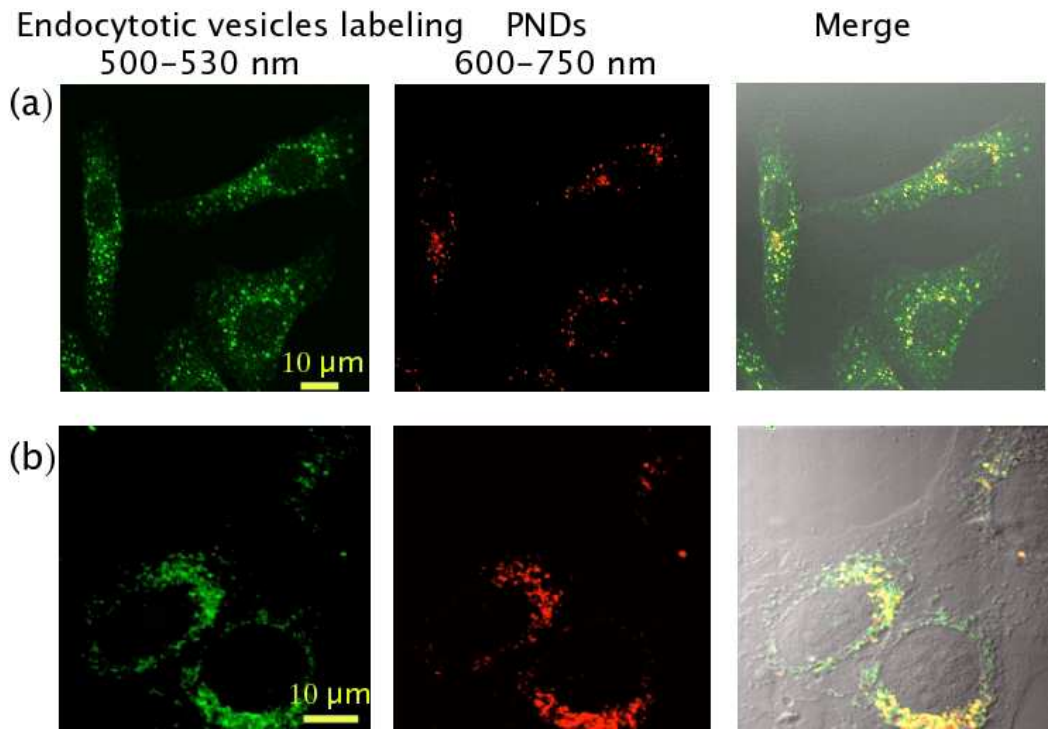}
\caption{Localization of 46~nm photoluminescent nanodiamonds in HeLa cells. Confocal fluorescence raster-scan (Leica TCS SP2 microscope) of HeLa cell incubated with PNDs (10~$\mu$g/ml) in normal (control) conditions, then fixed and labeled with endosomes or lysosomes dyes. From left to right: raster-scan in the \emph{green channel} (500-530~nm) showing the endocytic compartments; in the \emph{red channel} (600-750~nm) showing the PNDs. Images on the right represent the merged the green and red scans. (a) Colocalization study of PNDs with early endosomes labeled with EEA1-FITC fluorescent conjugate. (b) Colocalization of PNDs with lysosomes labeled with LysoTracker Green dye. PNDs colocalized with endosomes or lysosomes appear in yellow in the merged fluorescence scans.}
\label{fig:figure3}
\end{center}
\end{figure}

Early endosomes and lysosomes were marked with fluorescent labels emitting in the green spectral region (500-550~nm), with no overlap with the red and near infrared emission of NV color centers~({Figure~\ref{fig:figure5}d}). Scanning confocal imaging was carried out with the Leica TCS SP2 microscope. The \emph{red channel} detection spectral range selected for PNDs imaging is 600-750~nm to avoid again any overlap with the emission of the cellular components labels.
{Figure~\ref{fig:figure3}} shows a high degree of colocalization of PNDs with both early endosomes and lysosomes, which supports the fact that NDs follow the course of the endocytic cycle.
This endosomal localization of PNDs again agrees with reports on other kind of similar size nanoparticles, like QDs~\cite{Mattoussi-intracell} or gold nanobeads~\cite{Chithrani06, Brust08}.

\begin{figure}[ht!]
\begin{center}
\includegraphics[width=0.8\textwidth]{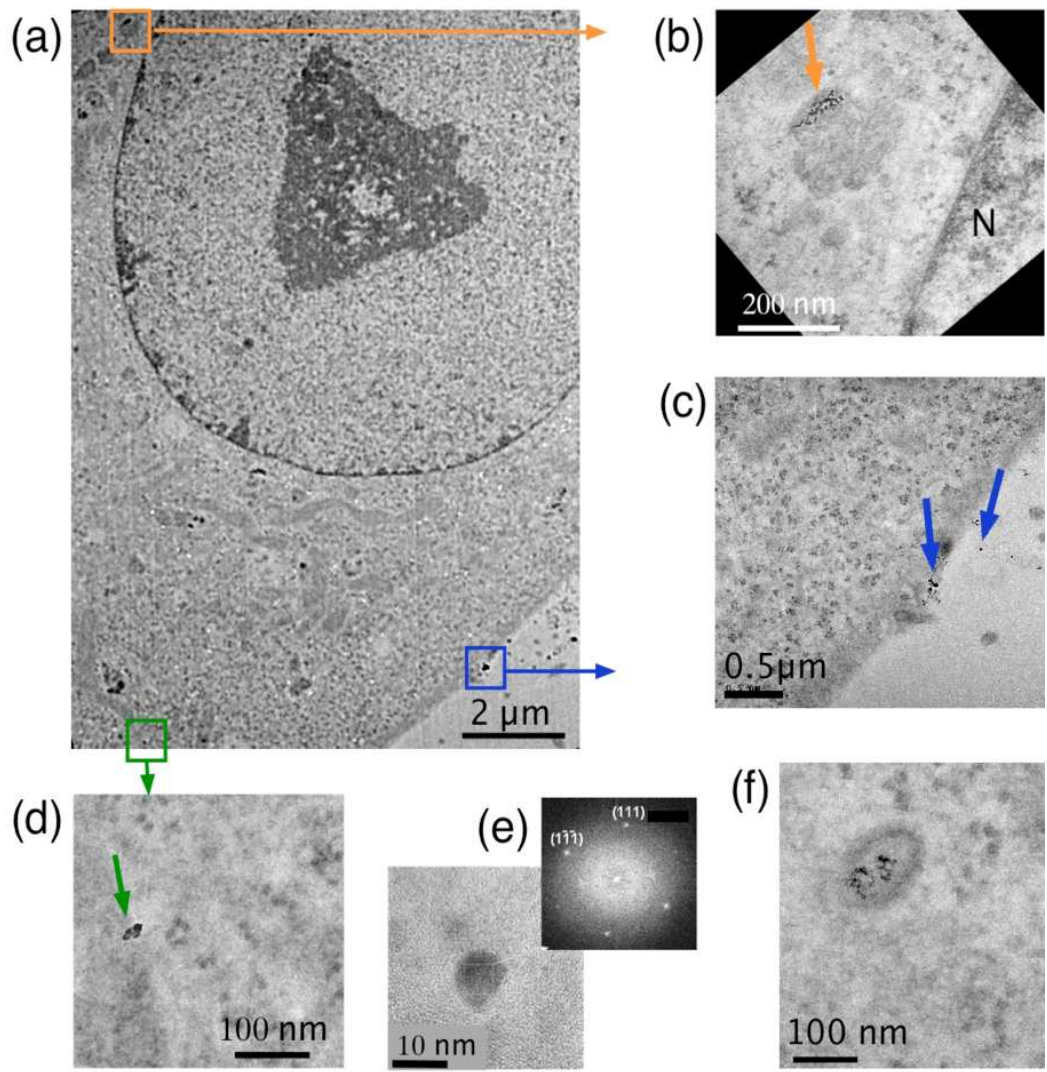}
\caption{Transmission Electron Microscopy images of a HeLa cell incubated with PNDs for 2~hours. (a) Large scale image. (b) NDs trapped in intracellular vesicles, N indicates the position of cell nucleus. (c) NDs outside the cell and on the cell membrane (blue arrows), probably during the early stage of the internalization process, since the half-life uptake (2.6~hours) is longer than the incubation time. (d) free NDs in the cytoplasm. (e) zoom on one 10~nm ND, inset : local area Fourier transform diffractogram of this ND. (f) PNDs trapped in an intracellular vesicle of another cell from the same sample.}
\label{fig:figure4}
\end{center}
\end{figure}

To further elucidate the fate of the smallest and less bright PNDs that cannot be detected with the Leica TCS SP2 microscope, we examined the same samples with two complementary techniques: with the home-built confocal setup detecting the smallest photoluminescent PNDs and with electron microscopy. 
Observations with the home-built confocal microscope show a partial colocalization, which is also confirmed by Transmission Electron Microscopy (TEM) imaging. In the latter technique, NDs particles appear as dark spots on the grey cytoplasmic background~\footnote{We examined by carrying out a local area Fourier transform electron diffractogram on one of the TEM dark spots of size $\simeq$10~nm that it has the signature of diamond material (inset of {Figure~\ref{fig:figure4}e}).}.

In the large scale TEM image ({Figure~\ref{fig:figure4}a}) of a part of a HeLa cell incubated with PNDs, one can observe the nanoparticles in the cytoplasm. They are either trapped in vesicles ({Figure~\ref{fig:figure4}b,f}), or free in the cytosol ({Figure~\ref{fig:figure4}d}). NDs trapped in vesicles form aggregates ({Figure~\ref{fig:figure4}b}), and represent the majority of the nanodiamonds observed by TEM inside the cell cytoplasm. This is in agreement with our colocalization studies and with the confined motion of PNDs observed with wide-field microscopy~\cite{Faklaris_08,Chang_08}. Nanodiamonds which are free in the cytosol correspond to the smallest particles that can be observed at their primary size (<5-10~nm) or as small aggregates of a  few particles. These free nanoparticles have either been released from the endosomes or may have been directly internalized via passive transport (like facilitated diffusion) through the cell membrane. Interestingly, all internalized nanodiamonds, even of the smallest size (<5~nm) are observed in perinuclear regions and none of them seems to be present inside the nucleus.

\section{Conclusion}
In this work we used 46~nm mean size photoluminescent diamond nanoparticles produced by micron-size diamond milling as cell labels. We showed that photoluminescence detection allows imaging the smallest nanodiamonds that cannot be observed by the backscattered light of the excitation laser. This result underlines the superiority of photoluminescence for tiny nanodiamonds imaging in complex environments  such as the intracellular medium. 
With the use of a commercial laser scanning confocal microscope, we showed that  the internalization of PNDs in HeLa cells occurs through endocytosis and we have strong indications that it is receptor-mediated via the clathrin pathway. 
We observed that most of the nanodiamonds are localized in intracellular endocytic vesicles in the perinuclear region, except for a small portion, in particular the smallest particles, that appear to be free in the cytoplasm. 
As cell is an inhomogeneous medium, a careful understanding of the distribution of nanoparticles into the cytoplasm and the reasons of the variability of their fate represents a critical step towards the use of diamond nanoparticles for targetted biomolecule delivery, with a long-term intracellular tracking possibility through its photoluminescence.

\section{Materials and Methods}
\subsection{Production and characterization of Photoluminescent NanoDiamonds (PNDs)}
The PNDs production is described in details in Ref.~\cite{Curmi09}.
Briefly, the starting material was type Ib diamond synthetic micron size powder (Element Six, The Netherlands) with a specified size of 150-200~$\mu$m. NV centers were created by electron irradiation (irradiation dose $2\times10^{18}$~e$^-$/cm$^2$, beam energy 8~MeV) and subsequent annealing (800$^\circ$C, 2~h) in vacuum (pressure $\simeq10^{-8}$~mbar) . Diamond microcrystals were then reduced in size by nitrogen jet milling to obtain submicron size crystals (Hosokawa-Alpine, Germany). Further size reduction to nanoparticles was achieved by ball milling under argon (Fritsch, Germany). The milled powder was then sieved and treated by strong sonication (60~min at 50°C), followed by acid treatment. After rinsing with pure water, filtration and centrifugation, we obtained PNDs with an average hydrodynamic diameter of 46~nm  in a stable water suspension. The size measurement was done by dynamic light Scattering (DLS) corrected from Mie scattering, using the DL135 particle size analyzer (Cordouan Technologies, France). The associated zeta potential of this solution was -43~mV (Zetasizer Nano ZS from Malvern, UK). 

\begin{figure}[ht!]
\begin{center}
\includegraphics[width=0.9\textwidth]{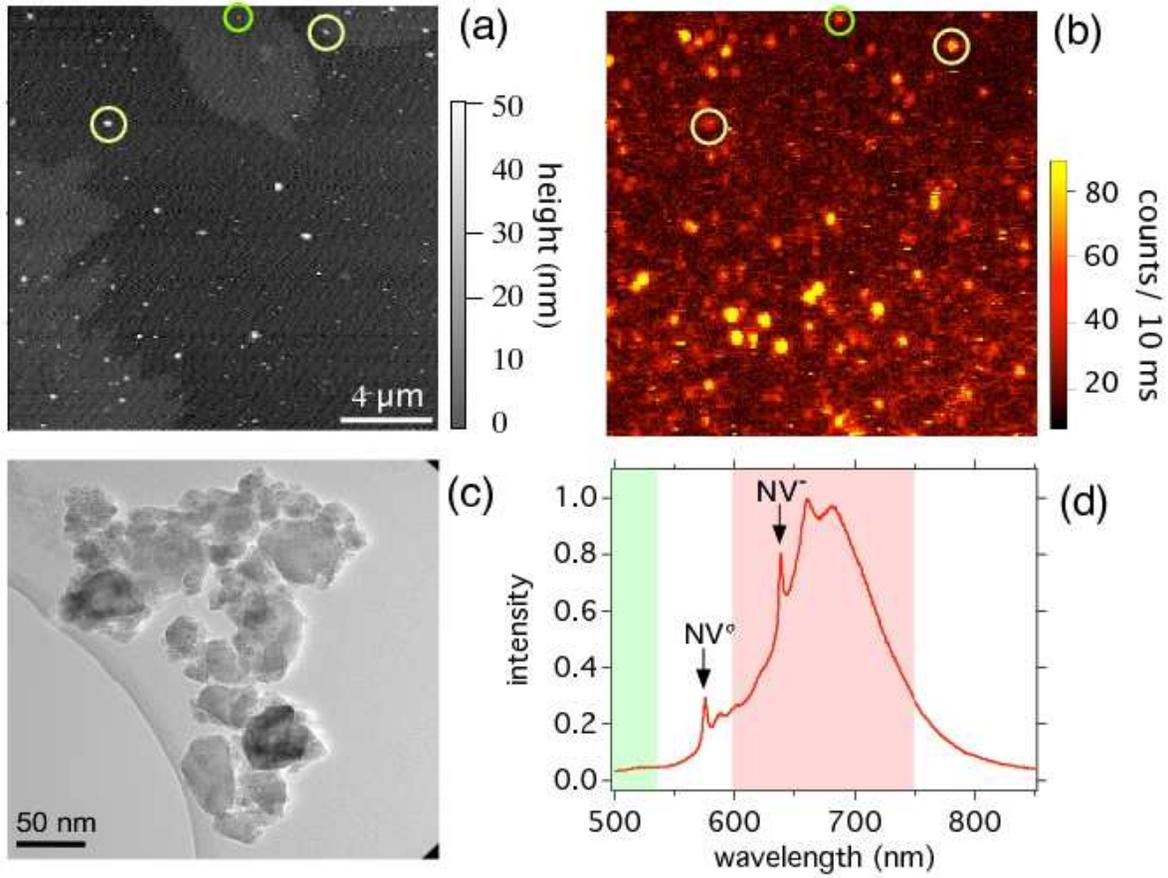}
\caption{Simultaneous size and photoluminescence characterization of the PNDs used in this work. Sample is a coverglass on which an aqueous suspension of PNDs is deposited by spin-coating. (a) and (b): raster scans ($18\times 18~\mu$m) of the same region of the sample in the two modes (a) AFM  and (b) confocal microscopy. For confocal imaging, the excitation laser wavelength is 532~nm, at an excitation power of 200~$\mu$W, and the photoluminescence intensity scale bar maximum of graph (b) was set to 90~counts/10 ms in order to be able to observe the less intense PNDs (the brightest spot yields 367~counts/10 ms). Two nanodiamonds of height  $\sim$40~nm are circled in yellow, while a 12~nm height nanodiamond is inside the green circle, still displaying a strong photoluminescence signal. (c) High-Resolution TEM image of the same type of PNDs. A droplet of the aqueous suspension was initially dried on the carbon grid. (d) Photoluminescence spectrum of NV color centers embedded in the starting micron-size diamond crystal (after irradiation and before milling). The narrow lines pointed out by arrows at 575~nm and 637~nm indicate respectively the zero-phonon lines of the neutral NV$^\circ$ and negatively charged NV$^-$ color centers, present in this sample. The light red/green rectangles behind NV spectra indicates the spectral range selected by the \emph{red/green channels} filters respectively on the Leica TCS SP2 microscope for PNDs .}
\label{fig:figure5}
\end{center}
\end{figure}

To simultaneously characterize the PNDs in size and photoluminescence intensity, we used an Atomic Force Microscope (AFM, MFP-3D Asylum Research) coupled to a home-built confocal microscope setup.
For such a study, a droplet of the PND aqueous suspension was deposited by spin-coating on a glass coverslip which was then simultaneously imaged in AFM ({Figure~\ref{fig:figure5}a}) and photoluminescence ({Figure~\ref{fig:figure5}b}) modes. Among 200 nanodiamonds identified on the AFM scan, 115 display a photoluminescence signal, i.e. about 58\% of the NDs are photoluminescent. Among the photoluminescent nanoparticles a small fraction consists of 10-15~nm diameter PNDs, but the majority of the PNDs appears as nano-objects of size 40-50~nm. Some of them form aggregates of 2-3 smaller particles, as can be inferred from TEM imaging ({Figure~\ref{fig:figure5}c}). 

\subsection{Photoluminescence and backscatterred light imaging of nanodiamonds}
The large scale photoluminescence observations of PNDs in the cells which are presented in the main text are done with a commercial Leica TCS SP2 (Manheim, Germany) laser scanning confocal microscope, with a $\times$63, 1.4 numerical aperture (NA) oil-immersion objective. Excitation comes from the cw argon ion laser line at 488~nm wavelength.  We use one Airy Unit as the pinhole diameter, for all the acquisitions. The detection was done in two non-overlapping spectral channels: \emph{green channel} (500-530~nm) collecting the fluorescence from FITC and LysoTracker-green dye conjugates, and the red channel (600-750~nm) collecting mainly NV$^-$ photoluminescence (see {Figure~\ref{fig:figure5}d}).

For the single particle analysis presented in the Supporting Information, the photoluminescence was probed using a home-built scanning-stage confocal microscope. It relies on a Nikon TE300 microscope, converted to a confocal setup with the appropriate modifications. The microscope objective used is a $\times$60, 1.4 numerical aperture (NA)  apochromatic oil immersion objective. The excitation is a cw laser emitting at the wavelength of 488~nm. The photoluminescence signal was acquired by an avalanche photodiode in the single-photon counting mode. 
This home-built confocal setup is sensitive to the photoluminescence of a single NV color center in a PND, in a cultured cell environment~\cite{Faklaris_08}. 
To record the backscattered excitation laser light, we used a properly oriented quarter-wave plate and polarizing beam-splitter cube assembly to redirect all the reflected light towards an other avalanche photodiode in the single-photon counting mode.

For the PNDs uptake kinetics study, we estimated from the confocal raster scans the mean fluorescence intensity per cell in the red channel. For this purpose we used the \emph{mean intensity} analysis/measure tool of $\emph{ImageJ}$ software (NIH, USA). About 40 cells were analyzed for each different incubation time.

\subsection{Cell culture conditions and test of nanodiamond cytotoxicity}
HeLa cells were grown in standard conditions on glass coverslips in Dulbecco modified Eagle medium  (DMEM) supplemented with 10\% fetal calf serum (FCS) and 1\% penicillin/streptomycin. 
To study the cellular uptake of PNDs, cells were seeded at a density of $2\times10^5$~cells/1.3~cm$^2$ and grown at 37$^\circ$C in a humidified incubator under 5\% CO$_2$ atmosphere. 24~h after seeding, the PNDs aqueous suspension was added to the cell culture medium. The cells were grown under similar conditions for an additional period of time (from 2~h for fluorescence examination, up to 24~h for the cytotoxicity tests). After incubation, the excess of PNDs was removed by washing the cells with phosphate buffer saline (PBS). The cells were then fixed with 4\% paraformaldehyde in PBS and mounted on microscope slides for phase contrast and confocal microscopy studies. 
A necessary requirement to use PNDs as biomolecule markers is their low cytotoxicity. As cytotoxicity depends on the cellular type used and the size, shape and charge of the nanoparticles~\cite{Drezek08}, we performed cytotoxicity tests for the PNDs used here. We show that they are not toxic \emph{in vitro} after 24~h of incubation with cells (data not shown).

\subsection{Cellular uptake of nanodiamonds}
To investigate the {\bf uptake mechanism of PNDs}, cells were treated as follows:\\
\emph{Experiments blocking the endocytosis (energy dependent process):}
for low temperature incubation, cells were grown as described above with the cell culture kept at 4$^\circ$C instead of 37$^\circ$C.
For the incubation with PNDs under ATP depletion, the cells were preincubated in PBS buffer, supplemented with 10~mM NaN$_3$ during 30~min at 37$^\circ$C and then PNDs were added.

For the {\bf investigation of the type of receptor-mediated endocytosis mechanism}, cells were treated as follows:\\
\emph{Hypertonic treatment to hinder the clathrin-mediated process}: the cells were pre-incubated for 30~min in PBS buffer supplemented with 0.45~M sucrose followed by incubation with PNDs at 37$^\circ$C.\\
\emph{Filipin treatment blocking caveolae pathway}: the cells were pretreated in PBS buffer, supplemented with filipin (5~$\mu$g/ml) for 30~min before exposure to PNDs at 37$^\circ$C. 

For {\bf PNDs intracellular localization analyses} by immunofluorescence, the endosomes were labeled with FITC-conjugated Mouse Anti-human Early Endosome Antigen EEA1 (Ref. 612006, BD Transduction Laboratories, USA)~\cite{Faklaris_08}. FITC dye has absorption/emission maxima at 490/520~nm respectively.
The lysosomes were labeled with LysoTracker Green DND-26 dye (L7526, Invitrogen,  USA), with absorption/emission maxima at 504/511~nm respectively. After two hours of incubation of PNDs at 37$^\circ$C with the cells, the medium was replaced with prewarmed new medium containing the LysoTracker probe (75~nM), for one additional hour of incubation. The medium was then replaced with fresh medium just before cell fixation as described above.

\subsection{Transmission electron microscopy}
The electron microscopy observations were done with a High Resolution Transmission Electron Microscope (HR-TEM, Tecnai F20 operating at 200~keV). Cells were seeded for 24~h in standard conditions (conditions similar to those used for fluorescence experiments). PNDs were added in cell culture medium and incubated for 2~h at 37$^\circ$C. Cells were then fixed in a solution of paraformaldehyde, glutaraldehyde and phosphate buffer for 45 min at room temperature. After dehydration with a graded series of ethanol, the cells were embedded in EPON resin. Ultrathin sections of the resin block were then cut (100~nm thickness) and stained with 2\% uranyl acetate, for a higher contrast imaging under the HR-TEM microscope.

\section*{Acknowledgements}

\footnotesize

We are grateful to Jean-Fran\c cois Roch and Karen Perronet for fruitful discussions. We thank G\'eraldine Dantelle for the measurements of diamond colloidal suspensions zeta potentials, Abdallah Slablab for the simultaneous measurement of the size and photoluminescence intensity of the nanodiamonds and Anne Tarrade for the preparation of the samples observed with Transmission Electron Microscopy. This work was supported by the European Commission through the project ``Nano4Drugs" (contract LSHB-2005-CT-019102), by Agence Nationale de la Recherche through the project ``NaDia" (contract ANR-2007-PNANO-045) and by a ``Ile-de-France'' Region \emph{C'Nano} grant under the project ``Biodiam".


\bibliography{Faklaris_et_al_final}
\bibliographystyle{unsrt}

\end{document}